\documentclass[a4paper]{amsart}
\usepackage[margin=2.5cm,papersize={19.75cm,28cm}]{geometry}
\usepackage{amssymb}           
\usepackage{hyperref}
\usepackage{eucal}

\numberwithin{equation}{section}

\newtheorem{definition}{Definition}[section]

\newtheorem{remarkth}[definition]{Remark}

\renewcommand{\emph}[1]{{\bfseries\itshape{#1}}}

\newcommand{\R}{\mathbb{R}}      
\newcommand{\N}{\mathbb{N}}      
\newcommand{\Z}{\mathbb{Z}}      
\newcommand{\F}{\mathbb{F}}

\newcommand\map[3]{#1\ \colon\ #2\longrightarrow#3}

\newcommand{\FL}{{Leg}_L}
\makeatletter
\newcommand\prol{\@ifstar{\@proldf}{\@prolpf}}  
\def\@prolpf{\@ifnextchar[{\@prolpf@wrt}{\@prolpf@}}
\def\@prolpf@wrt[#1]#2{\@ifnextchar[{\@prolpf@wrt@at{#1}{#2}}{\@prolpf@wrt@{#1}{#2}}}
\def\@prolpf@wrt@at#1#2[#3]{\prolsymbol^{#1}_{#3}#2}
\def\@prolpf@wrt@#1#2{\prolsymbol^{#1}#2}
\def\@prolpf@#1{\@ifnextchar[{\@prolpf@at{#1}}{\@prolpf@@{#1}}}
\def\@prolpf@at#1[#2]{\prolsymbol_{#2}#1}
\def\@prolpf@@#1{\prolsymbol#1}
\def\@proldf{\@ifnextchar[{\@proldf@wrt}{\@proldf@}}
\def\@proldf@wrt[#1]#2{\@ifnextchar[{\@proldf@wrt@at{#1}{#2}}{\@proldf@wrt@{#1}{#2}}}
\def\@proldf@wrt@at#1#2[#3]{\prolsymbol^{*#1}_{#3}#2}
\def\@proldf@wrt@#1#2{\prolsymbol^{*#1}#2}
\def\@proldf@#1{\@ifnextchar[{\@proldf@at{#1}}{\@proldf@@{#1}}}
\def\@proldf@at#1[#2]{\prolsymbol^*_{#2}#1}
\def\@proldf@@#1{\prolsymbol^*#1}
\def\prolsymbol{\mathcal{T}}
\makeatother

\newcommand{\pder}[2]{\frac{\partial #1}{\partial #2}}

\begin{document}

\title[Semi-discrete variational Mechanics]{Some applications of semi-discrete variational integrators to classical
field theories}

\author[M. de Le\'on]{M. de Le\'on}
\address{M. de Le\'on:
Instituto de Matem\'aticas y F{\'\i}sica
Fundamental, Consejo Superior de Investigaciones
Cient\'{\i}ficas, Serrano 123, 28006 Madrid,
Spain} \email{mdeleon@imaff.cfmac.csic.es}

\author[J.\ C.\ Marrero]{Juan C.\ Marrero}
\address{Juan C.\ Marrero:
Departamento de Matem\'atica Fundamental,
Facultad de Ma\-te\-m\'a\-ti\-cas, Universidad de
la Laguna, La Laguna, Tenerife, Canary Islands,
Spain} \email{jcmarrer@ull.es}

\author[D.\ Mart\'{\i}n de Diego]{David Mart\'{\i}n de Diego}
\address{D.\ Mart\'{\i}n de Diego:
Instituto de Matem\'aticas y F{\'\i}sica
Fundamental, Consejo Superior de Investigaciones
Cient\'{\i}ficas, Serrano 123, 28006 Madrid,
Spain} \email{d.martin@imaff.cfmac.csic.es}

\thanks{This work has been partially supported by MICYT (Spain)
Grants BMF 2003-01319, MTM 2004-7832 and  S-0505/ESP/0158 of the
CAM}

\keywords{Discrete mechanics, Classical Field
theories, nonlinear wave equation, Lagrangian
Mechanics, Hamiltonian Mechanics.}

\begin{abstract} We develop a semi-discrete version
of discrete variational mechanics with
applications to numerical integration of
classical field theories. The geometric
preservation properties are studied.
\end{abstract}


\maketitle

\section{Introduction}

The calculus of variations is a fundamental tool
in the description and understanding of Classical
Mechanics and it is an area of active research.
One part of this activity was dedicated to
uncover the geometrical structures behind such
formalisms. Many physical systems not only evolve
in time, as in Classical Mechanics,  but also
posses a continuous spatial structure. This is
the setting of  Classical Field Theories in both
the Lagrangian and Hamiltonian formalisms
\cite{gimmsy1,gimmsy2}.

  One traditional way to analyze these problems
has been to pass to the Hamiltonian formalism (or Lagrangian
formalism) using a space-time decomposition of the parameter
space, and then applying classical  Dirac's theory of constraints
(in the singular case, which is typical in field theories). In
this way, we obtain a well-posed system of equations of motion
that can be eventually integrated or numerically simulated. It is
clear that using this space-time decomposition we broke the
original covariance of the theory and perhaps some geometrical
structure is lost, but the treatment of the equations is, in many
aspects, more easy and some of the geometrical structure is still
preserved. Moreover, for numerical simulation of the equations of
motion, after the space-time decomposition, we eventually obtain a
Hamiltonian system and symplectic integration methods may be
useful to solve numerically the initial problem.

In this sense, it may be  useful to introduce  geometric
integrators, that is, numerical schemes which preserve some of the
extra features of geometric nature of the dynamical systems.
Usually, these integrators can run, in simulations, for long time
with lower spurious  effects (for instance, bad energy behavior
for conservative systems) than the traditional ones
\cite{Hair,Sanz}.

A particular case of geometric integrators are variational
integrators. These integrators have their roots in the optimal
control literature  in the 1960's and they enter in the
``geometric differential arena" after the pioneering work of
Veselov \cite{Vese1} and Moser and Veselov \cite{Mose,Vese1}. In
these papers, there appears the discrete action sum, discrete
Euler-Lagrange equations,  discrete Noether theorem... These
integrators have been adapted for Field
Theories\cite{Jaro2,mars2}. All these integrators have
demonstrated exceptionally good longtime behavior.

In this paper, we will develop the theory of
semi-discrete variational integrators for
Classical Field Theories.  The basic idea is to
consider an spatial truncation that reduces the
partial differential equations derived from the
Euler-Lagrange equations to a system of ordinary
differential equations \cite{LeRe,OlWeWu}. The
main objective of the paper is to study the
geometric properties after this spatial
truncation: preservation of forms, energy
preservation and momentum preservation.

\section{\sc Discrete variational calculus}
\label{Dvc}

 First, we will recall discrete variational calculus,
following the approach in \cite{mawest} and references therein. A
discrete Lagrangian is a map $L_d: Q \times Q\rightarrow \R$,
which  may be considered as an approximation of a continuous
Lagrangian $L: TQ\rightarrow \R$. Define the action sum $S_d:
Q^{N+1}\rightarrow \R$ corresponding to the Lagrangian $L_d$ by
\[
{S_d}(q_0, \ldots, q_N)=\sum_{k=1}^{N}
L_d(q_{k-1}, q_{k})\; ,
\]
where $q_k\in Q$ for $0\leq k\leq N$.

Observe that  for any covector $\alpha\in T_{(x_1,x_2)}^*(Q\times
Q)$, we have the decomposition $\alpha=\alpha_1+\alpha_2$ where
$\alpha_i\in T^*_{x_i} Q$, thus,
\[
dL_d(q_0, q_1)=D_{1} L_d(q_0, q_1)+D_{2} L_d(q_0, q_1)\; .
\]

The discrete variational principle   states that the solutions of
the discrete system determined by $L_d$ must extremize the action
sum given fixed points $q_0$ and $q_N$. Extremizing ${S_d}$ over
$q_k$, $1\leq k\leq N-1$, we obtain the following system of
difference equations
\begin{equation}\label{discreteeq}
 D_1L_d( q_k, q_{k+1})+D_2L_d( q_{k-1}, q_{k})=0\; .
\end{equation}
These  equations are usually called the {\em discrete
Euler-Lagrange equations}. Under some regularity hypothesis (the
matrix $(D_{12}L_d(q_k, q_{k+1}))$ is regular), it is possible to
define a (local) discrete flow $ \Upsilon: Q\times
Q\longrightarrow Q\times Q$, by $\Upsilon(q_{k-1}, q_k)=(q_k,
q_{k+1})$ from Equations (\ref{discreteeq}).

Define the  discrete Legendre transformations associated to  $L_d$
by
\[
\begin{array}{rrcl}
FL^{-}_d:& Q\times Q&\longrightarrow& T^*Q\\
     & (q_0, q_1)&\longmapsto & (q_0, -D_1 L_d(q_0, q_1))\; ,\\
     FL^+_d:& Q\times Q&\longrightarrow& T^*Q\\
     & (q_0, q_1)&\longmapsto & (q_1, D_2 L_d(q_0, q_1))\; ,
\end{array}
\]
and the 2-form
$\omega_d=(FL^+_d)^*\omega_Q=(FL^{-}_d)^*\omega_Q$,
where $\omega_Q$ is the canonical symplectic form
on $T^*Q$. The discrete algorithm determined by
$\Upsilon$ preserves the symplectic form
$\omega_d$, i.e., $\Upsilon^*\omega_d=\omega_d$.
Moreover, if the discrete Lagrangian is invariant
under the diagonal action of a Lie group $G$,
then the discrete momentum map $J_d: Q\times Q
\rightarrow {\frak g}^*$ defined by $ \langle
J_d(q_k, q_{k+1}), \xi\rangle=\langle D_2L_d(q_k,
q_{k+1}), \xi_Q(q_{k+1})\rangle $ is preserved by
the discrete flow. Therefore, these integrators
are symplectic-momentum preserving integrators.
Here, $\xi_Q$ denotes the fundamental vector
field determined by $\xi\in {\frak g}$, where
${\frak g}$ the Lie algebra of $G$.

\section{Variational Calculus in Classical Field Theories}

Consider a locally trivial fibration $\pi:
Y\longrightarrow X$, where $Y$ is an
$(m+n+1)$-dimensional manifold and $X$ is the
parameter space usually equipped with a global
decomposition in space and time, that is,
$X=\R\times P$, $\dim P=n$.  We shall also fix a
volume form on $X$, that will be denoted by
$\eta$. We can choose fibred coordinates
$(x^\mu,y^i)$ in $Y$, so that $\pi (x^{\mu}, y^i)
= (x^{\mu})$, where $(x^{\mu})=(x^0, x^1, \ldots,
x^n)$ and $x^0\equiv t$ represents the time
evolution. Assume that the volume form is $\eta =
dx^0\wedge \ldots\wedge dx^n $. Here, $0 \leq
\mu, \nu, ... \leq n$ and $1 \leq i, j, ... \leq
m$. We shall also use the following useful
notation $d^{n}x^\mu := \iota_{\partial/\partial
x^\mu}\eta$.

The first order jet prolongation $J^{1}\pi$ is
the manifold of classes $j^{1}_{x}\phi$ of
sections $\phi$ of $\pi$ (whose set will be
denoted by $\Gamma(\pi))$ around a point $x$ of
$X$ which have the same Taylor expansion up to
order one. $J^1\pi$ can be viewed as the
generalization of the phase space of the
velocities for Classical Mechanics. Therefore,
$J^{1}\pi$, which we shall denote by $Z$, is an
$(n+1+m+(n+1)m)$-dimensional manifold.  If we
have adapted coordinates $(x^{\mu}, y^i)$ in $Y$,
then we have induced coordinates in $Z$, given by
\[ (x^{\mu},y^{i}, z^i_{\mu})=(t, x^{1},\ldots,
x^n,y^{i}, z^i_t, z^i_{1}, \ldots, z^i_n).\]

Suppose that we are given a function $L: Z\longrightarrow \R$ of
class $C^2$ in its $(n+1+m+(n+1)m)$-arguments. Let $G_{ X}$ be a
compact ($n+1$)-dimensional submanifold on $X$. We can thus
construct the following functional
$$
{\mathcal J}_L(\phi)=\int_{G_{X}}L(x^{\mu},
y^i(x), y^i_{\mu}(x))\; dx^0\wedge \ldots \wedge
dx^n
$$
for all section $\phi\in \Gamma(\pi)$, where
$\phi(x)=(x^{\mu}, y^i(x), y^i_{\mu}(x))$

\begin{definition}
A section $\phi\in \Gamma(\pi)$ is a solution of the  variational
problem determined by $L$ if and only if $\phi$ is a critical
point of ${\mathcal J}_L$.
\end{definition}

Extremizing the functional ${\mathcal J}_L$ we obtain the
Euler-lagrange equations:
\[
\frac{\partial L}{\partial
y^i}-\frac{d}{dx^\mu}\left(\frac{\partial L}{
\partial z^i_\mu}\right)=0, \qquad 0\leq i \leq m
\]

Now, denote by $ \hat{p}_i^\mu :=
\pder{L}{z^i_\mu} $ and by $ \hat{p} := L -
z^i_\mu \hat{p}_i^\mu$, then for a given
Lagrangian function $L$ and a volume form $\eta$
we can construct  the \emph{Poincar\'e-Cartan
$(n+1)$-form}
\begin{eqnarray*}
 \Theta_L &=& \Big(L-z^i_{\mu}\frac{\partial L}{\partial
z^i_{\mu}}\Big)d^{n+1}x
+\pder{L}{z^i_\mu}dy^i\wedge d^nx^{\mu}\\
&=& (\hat{p}dx^\mu+\hat{p}^\mu_idy^i)\wedge d^{n}x^{\mu}
\end{eqnarray*}

From this form, we can also define the
\emph{Poincar\'e-Cartan $(n+2)$-form}  as
$\Omega_L := -d\Theta_L.$  In induced coordinates
is expressed as follows
\begin{eqnarray*}
\Omega_L &=& -( d\hat{p}\wedge dx^\mu + d\hat{p}^\mu_i\wedge
dy^i)\wedge d^{n}x^{\mu}
\end{eqnarray*}

>From these equations it is easy to derive an \emph{intrinsic
version of Euler-Lagrange equations}. In fact, a section $\phi \in
\Gamma (\pi)$ is an extremal of ${\mathcal J}_L$ if and only if
$$
(j^1\phi)^*(\iota_\xi\Omega_L)=0
$$
for every vector field $\xi$ on $Z$.

We refer the reader to reference \cite{gimmsy1} for more details
and a complete derivation of the equations and the geometric
framework for classical field theories.

\section{Motivating Example: The nonlinear wave equation}

Consider the nonlinear wave equation given by
\begin{equation}\label{aqe}
 u_{tt}=\delta_x\sigma'(u_x)-f'(u)
\end{equation}
where $u: U\subset\R^2\rightarrow \R$ and
$\sigma, f$ are smooth functions. If
$\sigma(u_x)=u_x^2/2$ then we obtain the
semi-linear wave equation.

Equation (\ref{aqe}) corresponds to the Euler-Lagrange equation
for the lagrangian function
\[
 L(u, u_t, u_x)= \frac{1}{2}
u_t^2-\sigma(u_x)-f(u)
\]
Observe that, in this particular case, the Lagrangian does not
depend on the parameter space $X=\R\times \R$.

Now, replace the $x$-derivative in the Lagrangian
by a simple difference (for simplicity, we will
work with a uniform grid of $N+1$ points,
$h=L/N$) as follows:
\[
 L_{sd}(u_0, u_1, (u_0)_t, (u_1)_t)= \frac{1}{2}
\left(\frac{(u_0)_t+(u_1)_t}{2}\right)^2
-\sigma\left(\frac{u_1-u_0}{h}\right)-f\left(\frac{u_1+u_0}{2}\right)
\]

In this case, the Lagrangian is a function $L_{sd}: T\R\times
T\R\simeq T(\R\times \R)\longrightarrow \R$.  This
spatial-discretization is useful, since as we will show in the
next section, applying a suitable variational principle we obtain
a semi-discretization of Equation (\ref{aqe}), replacing the
partial differential equation by a system of ordinary differential
equations. After this step, the equations of motion can be
integrated in time or numerically integrated using, for instance,
a symplectic method as the symplectic Euler scheme.

In the next sections, we will show that with
these semi-discretizations
 some of the geometric properties are preserved.

\section{Semi-discrete variational calculus}
Given a smooth manifold $Q$ consider the following
sets
\begin{eqnarray*}
 {\mathcal C}_{N, [0,T]}&=&\left\{ y: \{0,\ldots, N\}\times [0,
T]\longrightarrow Q\; \Big| \; y(k, \cdot)\in C^2 ([0, T]) \hbox{
for all } k\right. \\&& \left. \hbox{ and the curves } t\to y(0,t)
\hbox{
and } t\to y(N, t) \hbox{ are fixed}\right\},\\
 \widetilde{\mathcal C}_{N, [0,T]}&=&\left\{ y: \{0,\ldots, N\}\times [0,
T]\longrightarrow Q\; \Big|\;  y(k, \cdot)\in C^2 ([0, T]) \right.
\\&& \left. \hbox{ and the values }  y(k,0) \hbox{ and } y(k, T)
\hbox{ are fixed for all } k \right\}.
\end{eqnarray*}
The choice of one of these sets depends of the different boundary
conditions of the initial problem. Of course, another boundary
conditions can be analyzed considering suitable adaptations of
both situations.

Now, let $L: TQ \times TQ \simeq T(Q \times Q) \to \R$ be a
Lagrangian function. If $(v_{0}, v_{1}) \in T_{q_{0}}Q \times
T_{q_{1}}Q$ then, as in Section \ref{Dvc}, we have that
\[
dL(v_{0}, v_{1}) = (\tilde{D}_{1}L)(v_{0}, v_{1}) +
(\tilde{D}_{2}L)(v_{0}, v_{1}),
\]
with $(\tilde{D}_{1}L)(v_{0}, v_{1}) \in T_{v_{0}}^{*}(TQ)$ and
$(\tilde{D}_{2}L)(v_{0}, v_{1}) \in T_{v_{1}}^{*}(TQ)$.

On the other hand, if $(q^{i}_{0})$ (respectively, $(q^{i}_{1})$)
are local coordinates on an open subset $U_{0}$ (respectively,
$(U_{1})$) of $Q$ such that $q_{0} \in U_{0}$ (respectively,
$q_{1} \in U_{1}$), then we may consider the corresponding local
coordinates $(q^{i}_{0}, \dot{q}^{i}_{0}, q^{i}_{1},
\dot{q}^{i}_{1})$ on $TQ \times TQ$ and it follows that
\[
\begin{array}{rcl}
\tilde{D}_{1}L (v_{0}, v_{1})& =& \displaystyle
\sum_{i=1}^{n}(\displaystyle\frac{\partial L}{\partial
q^{i}_{0}}_{|(v_{0}, v_{1})} dq^{i}_{0}(v_{0}, v_{1}) +
\displaystyle\frac{\partial L}{\partial \dot{q}^{i}_{0}}_{|(v_{0},
v_{1})} d\dot{q}^{i}_{0}(v_{0},
v_{1})),\\[8pt]
\tilde{D}_{2}L (v_{0}, v_{1}) &=& \displaystyle
\sum_{i=1}^{n}(\displaystyle\frac{\partial L}{\partial
q^{i}_{1}}_{|(v_{0}, v_{1})} dq^{i}_{1}(v_{0}, v_{1}) +
\displaystyle\frac{\partial L}{\partial \dot{q}^{i}_{1}}_{|(v_{0},
v_{1})} d\dot{q}^{i}_{1}(v_{0}, v_{1})).
\end{array}
\]

Moreover, we will use the following notation
\[
\begin{array}{rcl}
D_{1}L(v_{0}, v_{1}) &=& \displaystyle \sum_{i=1}^{n}
\frac{\partial L}{\partial q^{i}_{0}}_{|(v_{0}, v_{1})}
dq^{i}_{0}(v_{0}, v_{1}),\\[8pt] D_{2}L(v_{0}, v_{1}) &=& \displaystyle
\sum_{i=1}^{n} \frac{\partial L}{\partial
\dot{q}^{i}_{0}}_{|(v_{0}, v_{1})} d\dot{q}^{i}_{0}(v_{0}, v_{1}),
\\[8pt]
D_{3}L(v_{0}, v_{1}) &=& \displaystyle \sum_{i=1}^{n}
\frac{\partial L}{\partial q^{i}_{1}}_{|(v_{0}, v_{1})}
dq^{i}_{1}(v_{0}, v_{1}),\\[8pt] D_{4}L(v_{0}, v_{1}) &=& \displaystyle
\sum_{i=1}^{n} \frac{\partial L}{\partial
\dot{q}^{i}_{1}}_{|(v_{0}, v_{1})} d\dot{q}^{i}_{1}(v_{0}, v_{1}).
\end{array}
\]

Note that $D_{1}L(v_{0}, v_{1})$ and $D_{3}L(v_{0}, v_{1})$ may be
considered as $1$-forms on $Q$ at the points $q_{0}$ and $q_{1}$,
respectively. In fact,
\[
\begin{array}{rcl}
((\tau_{Q})_{*}^{v_{0}})^{t}(\displaystyle \sum_{i=1}^{n}
\frac{\partial L}{\partial q^{i}_{0}}_{|(v_{0}, v_{1})}
dq^{i}_{0}(q_{0}))& =& D_{1}L (v_{0}, v_{1}),\\[8pt]
((\tau_{Q})_{*}^{v_{1}})^{t}(\displaystyle \sum_{i=1}^{n}
\frac{\partial L}{\partial q^{i}_{1}}_{|(v_{0}, v_{1})}
dq^{i}_{1}(q_{1})) &=& D_{3}L (v_{0}, v_{1}),\end{array}
\]

\noindent where $((\tau_{Q})_{*}^{v_{0}})^{t}: T_{q_{0}}^{*}Q \to
T_{v_{0}}^{*}(TQ)$ (respectively, $((\tau_{Q})_{*}^{v_{1}})^{t}:
T_{q_{1}}^{*}Q \to T_{v_{1}}^{*}(TQ)$) is the dual map of the
linear epimorphism $(\tau_{Q})_{*}^{v_{0}}: T_{v_{0}}(TQ) \to
T_{q_{0}}Q$ (respectively, $(\tau_{Q})_{*}^{v_{1}}: T_{v_{1}}(TQ)
\to T_{q_{1}}Q$). In addition, using the canonical identification
between the vector spaces $T_{q_{0}}Q$ and $ker
(\tau_{Q})_{*}^{v_{0}} = < \displaystyle \frac{\partial}{\partial
\dot{q}^{i}_{0}}_{|v_{0}} >$ (respectively, $T_{q_{1}}Q$ and $ker
(\tau_{Q})_{*}^{v_{1}} = < \displaystyle \frac{\partial}{\partial
\dot{q}^{i}_{1}}_{|v_{1}} >$) we have that the $1$-form
$D_{2}L(v_{0}, v_{1})$ (respectively, $D_{4}L(v_{0}, v_{1})$) may
be considered as the $1$-form on $Q$ at the point $q_{0}$
(respectively, $q_{1}$)
\[
\displaystyle \sum_{i=1}^{n} \frac{\partial L}{\partial
\dot{q}^i_0}_{|(v_{0}, v_{1})} dq^i_{0}(q_{0})
\]
(respectively, $\displaystyle \sum_{i=1}^{n}
\frac{\partial L}{\partial \dot{q}^i_1}_{|(v_{0},
v_{1})} dq^{i}_{1}(q_{1})$).

\subsection{Variational calculus on ${\mathcal C}_{N, [0,T]}$}
Define the \emph{semi-discrete action} ${\mathcal S}_{\hbox{\small
sd}} L : {\mathcal C}_{N, [0,T]}\longrightarrow \R$ as follows
\[
\begin{array}{rcl} {\mathcal S}_{\hbox{\small sd}} L ( y(\cdot,
\cdot))&=& \int^T_0 \left[\sum_{k=0}^{N-1} L( \dot{y}(k, t),
\dot{y}(k+1, t))\right]\, dt\\[8pt]
&=&  \int^T_0 \left[\sum_{k=0}^{N-1} L( y(k,t), \dot{y}(k, t),
{y}(k+1, t), \dot{y}(k+1, t))\right]\, dt\; .
\end{array}
\]

\begin{definition}
An element $y\in {\mathcal C}_{N, [0,T]}$ is a solution of the
semi-discrete variational problem determined by $L$ if and only if
it is a critical point of the Lagrangian system defined by
${\mathcal S}_{\hbox{\small sd}} L$.
\end{definition}

Therefore, a solution $y$ of the semi-discrete variational problem
extremizes ${\mathcal S}_{\hbox{\small sd}} L$ among all the
possible variations of $y$, where a variation of $y$ is a smooth
curve $ s\in (-\epsilon, \epsilon)\longrightarrow y_s\in {\mathcal
C}_{N, [0,T]}$ with $y_0=y$. Denote by
\[
\delta y_k(t)=\displaystyle{\frac{d y_s}{ds}(k,t)}\Big|_{s=0}
\]
where we use the notation $y_k(t)=y(k, t)$.
Observe that $y_s(0,t)=y_0(t)$ and $y_s(N,
t)=y_N(t)$ for all $s$. We will also use the
following notation \begin{eqnarray*}
 D_{i}L_{(k,
k+1)}(t) &=& (D_{i}L)(y_{k}(t), \dot{y}_{k}(t),
y_{k+1}(t), \dot{y}_{k+1}(t))\; ,\\
\tilde{D}_{j}L_{(k, k+1)}(t) &=&
(\tilde{D}_{j}L)(y_{k}(t), \dot{y}_{k}(t),
y_{k+1}(t), \dot{y}_{k+1}(t))\; \end{eqnarray*}
for $i \in \{1, \dots , 4\}$ and $j \in \{1,
2\}$.

 Extremizing the semi-discrete action function among all the possible variations, we
find that
\begin{eqnarray*}
&&\displaystyle{\frac{d}{ds}}\Big|_{s=0}\left({\mathcal
S}_{\hbox{\small sd}} L (y_s)\right)= \frac{d}{ds}\Big|_{s=0}
\int^T_0 \left[\sum_{k=0}^{N-1} L( y_s(k,t), \dot{y_s}(k, t),
{y}_s(k+1, t), \dot{y_s}(k+1, t))\right]\, dt\\
&&= \int^T_0 \sum_{k=1}^{N-1} \left[ D_3 L_{(k-1, k)}(t){\mathbf
\delta} y_{k}(t)  + D_4 L_{(k-1, k)}(t)\frac{d}{dt}{\mathbf
\delta} y_{k}(t)\right. \\&&\left. +D_1 L_{(k, k+1)}(t){\mathbf
\delta} y_{k}(t) + D_2 L_{(k, k+1)}(t)\frac{d}{dt}{\mathbf \delta}
y_{k}(t)
\right]\, dt\\
&&= \int^T_0 \left[\sum_{k=1}^{N-1} \left(D_3 L_{(k-1, k)}(t) +D_1
L_{(k, k+1)}(t)- \frac{d}{dt} \left( D_4 L_{(k-1, k)}(t) + D_2
L_{(k, k+1)}(t)\right)\right){\mathbf \delta} y_{k}(t)
\right]\, dt   \\
&&         + \displaystyle \sum_{k=1}^{N-1} \left( D_4 L_{(k-1,
k)}(t) + D_2 L_{(k, k+1)}(t)\right){\mathbf \delta}
y_{k}(t)\Big|^T_0.
\end{eqnarray*}

Therefore, the \emph{semi-discrete Euler-Lagrange equations} are:
\begin{eqnarray}
D_3 L_{(k-1, k)}(t) +D_1 L_{(k, k+1)}(t)- \frac{d}{dt} \left( D_4
L_{(k-1, k)}(t) + D_2 L_{(k, k+1)}(t)\right)&=&0,\quad 1\leq k\leq
N-1 \nonumber\\
D_4 L_{(k-1, k)}(0) + D_2 L_{(k, k+1)}(0)&=& 0 \label{ele}\\
D_4 L_{(k-1, k)}(T) + D_2 L_{(k, k+1)}(T)&=& 0
\nonumber
\end{eqnarray}

The first equations represent a system of second
order differential equations of the form:
\[
F(y_{k-1}(t), y_{k}(t), y_{k+1}(t); \dot{y}_{k-1}(t),
\dot{y}_{k}(t), \dot{y}_{k+1}(t);  \ddot{y}_{k-1}(t),
\ddot{y}_{k}(t), \ddot{y}_{k+1}(t))=0
\]
When  the matrix $(D_{24}L_{(j-1, j)}(t))$ is regular then we may
locally write these equations as \begin{equation}\label{ppp}
 \ddot{y}_{k+1}(t)=G(y_{k-1}(t), y_{k}(t), y_{k+1}(t); \dot{y}_{k-1}(t),
\dot{y}_{k}(t), \dot{y}_{k+1}(t);  \ddot{y}_{k-1}(t),
\ddot{y}_{k}(t)) \end{equation}
 Then, for
enough small $T$, the \emph{semi-discrete flow} $\Upsilon$:
\[
\begin{array}{rrcl}
\Upsilon:& Q^{[0,T]}\times
Q^{[0,T]}&\longrightarrow &Q^{[0,T]}\times
Q^{[0,T]} \\
& (y_{k-1}(\cdot), y_{k}(\cdot))&\longmapsto &(y_{k}(\cdot),
y_{k+1}(\cdot)) \end{array}
\]
 is well-defined (since Equations (\ref{ppp}) appear as a system of explicit differential equations) , where $(y_{k-1}(\cdot), y_{k}(\cdot),
 y_{k+1}(\cdot))$ satisfies Equations (\ref{ele}) and
\[
Q^{[0, T]} = \{z: [0, T] \to Q / z \in C^{2}([0, T])\}.
\]

\subsubsection{Symplecticity}

Define the Poincar\'e-Cartan 1-forms $\Theta_{L}^-,
\Theta_{L}^+\in \Lambda^1(Q^{[0,T]}\times Q^{[0,T]})$ as follows
\[\begin{array}{rcl}
\Theta_{L}^-(X(\cdot), Y(\cdot)) &=& -\int_0^T\left(\tilde{D}_1
L_{(0, 1)}(t)\dot{X}(t)\right) dt\\[8pt]
&=& -\int_0^T\left(D_1 L_{(0, 1)}(t)X(t)+
 D_2 L_{(0, 1)}(t)\dot{X}(t)\right)\, dt \\[8pt]
&=& -\int_0^T\left(D_1 L_{(0, 1)}(t)- \frac{d}{dt} \left(  D_2
L_{(0, 1)}(t)\right)\right)X(t)\, dt - D_2 L_{(0, 1)}(t)
X(t)\Big|_{0}^T\\
\Theta_{L}^+(X(\cdot), Y(\cdot)) &=& \int_0^T\left(\tilde{D}_2
L_{(0, 1)}(t)\dot{Y}(t)\right)\, dt\\[8pt]
&=& \int_0^T\left(D_3 L_{(0, 1)}(t)Y(t)+
 D_4 L_{(0, 1)}(t)\dot{Y}(t)\right)\, dt\\[8pt]
&=& \int_0^T\left(D_3 L_{(0, 1)}(t)- \frac{d}{dt} \left(  D_4
L_{(0, 1)}(t)\right)\right)Y(t)\, dt + D_4 L_{(0, 1)}(t)
Y(t)\Big|_{0}^T\end{array}\]

Then, we have
\[
 \Theta_{L}^+(X(\cdot),
Y(\cdot))- \Theta_L^-(X(\cdot), Y(\cdot))=d\left[\int^T_0 L(
y_0(t), \dot{y}_0(t), {y}_1(t), \dot{y}_1(t))\,
dt\right](X(\cdot), Y(\cdot))
\]
Therefore, there exists a well-defined 2-form $\Omega_L$
\[
\Omega_L=d\Theta_{L}^-=d\Theta_{L}^+
\]
As a consequence of Equations (\ref{ele}) we
deduce that $\Upsilon^*\Theta^-_L=\Theta^+_L$ and
\[
\Upsilon^*\Omega_L=\Omega_L
\]

\subsubsection{Legendre transformations}

Define the semi-discrete Legendre
transformations as
\[
\begin{array}{rrclrcl}
 \FL^-:& Q^{[0,T]}\times Q^{[0,T]} &\longrightarrow &
 T^*Q^{[0,T]}& & &\\
&(y_{0}(\cdot), y_{1}(\cdot))&\longmapsto & \FL^-(y_{0}(\cdot),
y_{1}(\cdot)):& T_{y_0(\cdot)} Q^{[0,T]}& \longrightarrow&\R\\
 & & & & X(\cdot)&\longmapsto &\Theta_L^-(X(\cdot), \mathbf{0})
\end{array}
\]
and
\[
\begin{array}{rrclrcl}
 \FL^+:& Q^{[0,T]}\times Q^{[0,T]} &\longrightarrow &
 T^*Q^{[0,T]}& & &\\
&(y_{0}(\cdot), y_{1}(\cdot))&\longmapsto & \FL^+(y_{0}(\cdot),
y_{1}(\cdot)):& T_{y_1(\cdot)} Q^{[0,T]}& \longrightarrow&\R\\
 & & & & Y(\cdot)&\longmapsto &\Theta_L^+(\mathbf{0}, Y(\cdot))
\end{array}
\]
Denote by $\Theta_Q$ and $\Omega_Q$ the Liouville 1-form and the
canonical symplectic 2-form, respectively,  on $T^*Q^{[0,T]}$
defined by
\[
\begin{array}{rcl}
 \Theta_Q(\tilde{\alpha}(\cdot))(\tilde{X}(\cdot)) &=&
 \tilde{\alpha}(\cdot)((\tau_{Q^{[0, T]}})_{*}(\tilde{X}(\cdot)))
 =
\int^T_0
\tilde{\alpha}((\tau_Q)_*(\tilde{X}(t)))\; dt\\
\Omega_Q &=& d \Theta_Q
\end{array}
\]
where $\tau_Q: T^*Q\longrightarrow Q$ and $\tau_{Q^{[0, T]}}:
T^*Q^{[0, T]} \to Q^{[0, T]}$ are the canonical projections.
 Then, it is easy to prove that
 \begin{eqnarray*}
(\FL^-)^*\Theta_Q=\Theta_L^-\; \ , \quad
(\FL^-)^*\Theta_Q=\Theta_L^+, \quad  (\FL^-)^*\Omega_Q=\Omega_L.
\end{eqnarray*}

\subsubsection{Momentum mapping}\label{mp}

Suppose that the Lagrangian is invariant by a Lie group of
symmetries, that is, if $\phi: G\times Q\to Q$ is the action of a
Lie group then
\[
L(T_q \phi_g(v_q), T_{q'} \phi_g(v_{q'}))= L(v_q, v_{q'}), \quad
\forall v_q, v_{q'}\in TQ, \quad q, q'\in Q
\]
Infinitesimally, this condition means that
\[
\xi_{T(Q\times Q)}(v_q,
v_{q'})(L)=\left(\xi_{TQ}(v_q)+\xi_{TQ}(v_{q'})\right)(L)=0, \quad
\forall \xi\in{\frak g}
\]
where $\xi_{TQ}$ is the infinitesimal generator of the lifted
action $\phi^T: G\times TQ\to TQ$. In particular,
\[
\left(\xi_{TQ}(y_{k-1}(t),
\dot{y}_{k-1}(t))+\xi_{TQ}(y_{k}(t),
\dot{y}_{k}(t))\right)(L)=0
\]
or, in other words,
\begin{eqnarray*}
&&D_1 L_{(k-1, k)}(t)\xi_Q(y_{k-1}(t))+D_2 L_{(k-1,
k)}(t)\frac{d}{dt}\left(\xi_Q(y_{k-1}(t))\right)\\ &&+ D_3
L_{(k-1, k)}(t)\xi_Q(y_{k}(t))+D_4 L_{(k-1,
k)}(t)\frac{d}{dt}\left(\xi_Q(y_{k}(t))\right)=0,
\end{eqnarray*}
where $\xi_{Q}$ is the infinitesimal generator of the action
$\phi$ corresponding to $\xi$ (note that $\xi_{TQ}$ is the
complete lift of $\xi_{Q}$).  Therefore

\begin{eqnarray*}
 &&D_3 L_{(k-1, k)}(t)\xi_Q(y_{k}(t)) +D_1
L_{(k-1, k)}(t)\xi_Q(y_{k-1}(t))\\
&&- \frac{d}{dt} \left( D_4 L_{(k-1, k)}(t)\right)\xi_Q(y_{k}(t))
- \frac{d}{dt} \left(D_2 L_{(k-1,
k)}(t)\right)\xi_Q(y_{k-1}(t))\\
&&+\frac{d}{dt}\left[D_4 L_{(k-1, k)}(t)\xi_Q(y_{k}(t)) + D_2
L_{(k-1, k)}(t)\xi_{Q}( y_{k-1}(t))\right]=0
\end{eqnarray*}
 Subtracting  first expression in (\ref{ele})
applied to $\xi_{T(Q\times Q)}$ and the above equation we deduce
that
\begin{eqnarray*}
&&D_1 L_{(k, k+1)}(t)\xi_Q(y_{k}(t))- \frac{d}{dt} \left( D_2
L_{(k,k+1)}(t)\right)\xi_Q(y_{k}(t))\\
&=& D_1 L_{(k-1, k)}(t)\xi_Q(y_{k-1}(t))- \frac{d}{dt} \left( D_2
L_{(k-1,k)}(t)\right)\xi_Q(y_{k-1}(t))\\
&& +\frac{d}{dt}\left[D_4 L_{(k-1, k)}(t)\xi_Q(y_{k}(t)) + D_2
L_{(k-1, k)}(t)\xi_{Q}( y_{k-1}(t))\right].
\end{eqnarray*}

Therefore, integrating and using the two last equations in
(\ref{ele}) we obtain the following preservation law
\begin{eqnarray*}
 &&\int_0^T \left[D_1 L_{(k-1, k)}(t)- \frac{d}{dt} \left( D_2
L_{(k-1,k)}(t)\right)\right]\xi_Q(y_{k-1}(t))\, dt + D_{2}L_{(k-1,
k)}(t) \xi_{Q}(y_{k-1}(t))\Big|_{0}^T \\
&& = \int_0^T \left[D_1 L_{(k, k+1)}(t)- \frac{d}{dt} \left( D_2
L_{(k,k+1)}(t)\right)\right]\xi_Q(y_{k}(t))\, dt + D_{2}L_{(k,
k+1)}(t) \xi_{Q}(y_{k}(t))\Big|_{0}^T.
\end{eqnarray*}

Note that this equation may be written as
\[
\Theta_{L}^{-}(y_{k-1}(\cdot), y_{k}(\cdot))(\xi_{Q}\circ y_{k-1},
\xi_{Q} \circ y_{k}) = \Theta_{L}^{-}(y_{k}(\cdot),
y_{k+1}(\cdot))(\xi_{Q}\circ y_{k}, \xi_{Q} \circ y_{k+1}).
\]

\subsubsection{The nonlinear wave equation. First point of view}

The semi-discrete Euler-Lagrange equations for the Lagrangian:
\[
 L_{sd}(u_0, u_1, (u_0)_t, (u_1)_t)= \frac{1}{2}
\left(\frac{(u_0)_t+(u_1)_t}{2}\right)^2
-\sigma\left(\frac{u_1-u_0}{h}\right)-f\left(\frac{u_1+u_0}{2}\right)
\]

are: \begin{eqnarray*}
&&\frac{({u}_{k-1})_{tt}+2({u}_k)_{tt}+({u}_{k+1})_{tt}}{4}-\frac{1}{h}\left[\sigma'\left(\frac{u_{k+1}-u_k}{h}\right)
-\sigma'\left(\frac{u_{k}-u_{k-1}}{h}\right)\right]\\
&&-\frac{1}{2}\left[f'(\frac{u_k+u_{k-1}}{2})
+f'(\frac{u_{k+1}+u_{k}}{2})\right]=0,\quad 1\leq k\leq N-1
\end{eqnarray*}
with boundary conditions
\begin{eqnarray}
&&u_0(t) \hbox{ and } u_1(t) \hbox{ fixed}\nonumber\\
&& (u_{k-1})_t(0)+2(u_{k})_t(0)+(u_{k+1})_t(0)=0\label{de1}\\
&&
(u_{k-1})_t(T)+2(u_{k})_t(T)+(u_{k+1})_t(T)=0\label{de2}
\end{eqnarray}

\subsection{Variational calculus on $\widetilde{\mathcal C}_{N, [0,T]}$}

Instead of the set of functions  ${\mathcal C}_{N, [0,T]}$ we
consider the set $\widetilde{\mathcal C}_{N, [0,T]}$ where now the
values at time $0$ and $T$ are fixed. As in the previous case, we
have the
 \emph{semi-discrete action}
$\widetilde{\mathcal S}_{\hbox{\small sd}} L : \widetilde{\mathcal
C}_{N, [0,T]}\longrightarrow \R$:
\[
\widetilde{\mathcal S}_{\hbox{\small sd}} L ( y(\cdot,
\cdot))=\int^T_0 \left[\sum_{k=0}^{N-1} L( y(k,t), \dot{y}(k, t),
{y}(k+1, t), \dot{y}(k+1, t))\right]\, dt\; .
\]
\begin{definition}
A function $y\in \widetilde{\mathcal C}_{N, [0,T]}$ is a solution
of the semi-discrete variational problem determined by $L$ if and
only if it is a critical point of the Lagrangian system defined by
$\widetilde{\mathcal S}_{\hbox{\small sd}} L$. \end{definition}

 Therefore
\begin{eqnarray*}
&&\displaystyle{\frac{d}{ds}}\Big|_{s=0}\left(\widetilde{\mathcal
S}_{\hbox{\small sd}} L (y_s)\right)= \frac{d}{ds}\Big|_{s=0}
\int^T_0 \left[\sum_{k=0}^{N-1} L( y_s(k,t), \dot{y_s}(k, t),
{y}_s(k+1, t), \dot{y_s}(k+1, t))\right]\, dt\\
&&= \int^T_0 \sum_{k=1}^{N-1} \left[ D_3 L_{(k-1, k)}(t){\mathbf
\delta} y_{k}(t)  + D_4 L_{(k-1, k)}(t)\frac{d}{dt}{\mathbf
\delta} y_{k}(t)\right. \\&&\left. +D_1 L_{(k, k+1)}(t){\mathbf
\delta} y_{k}(t) + D_2 L_{(k, k+1)}(t)\frac{d}{dt}{\mathbf \delta}
y_{k}(t) \right]\, dt\\
&& + \int^T_0 \left[D_1 L_{(0, 1)}(t){\mathbf \delta} y_{0}(t)  +
D_2 L_{(0, 1)}(t)\frac{d}{dt}{\mathbf \delta}
y_{0}(t)\right]\, dt\\
&& + \int^T_0 \left[D_3 L_{(N-1, N)}(t){\mathbf \delta} y_{N}(t) +
D_4 L_{(N-1, N)}(t)\frac{d}{dt}{\mathbf \delta} y_{N}(t)\right]\,
dt
\\
&&= \int^T_0 \left[\sum_{k=1}^{N-1} \left(D_3 L_{(k-1, k)}(t) +D_1
L_{(k, k+1)}(t)- \frac{d}{dt} \left( D_4 L_{(k-1, k)}(t) + D_2
L_{(k, k+1)}(t)\right)\right){\mathbf \delta} y_{k}(t)
\right]\, dt   \\
&&+ \int^T_0 \left[D_1 L_{(0, 1)}(t)-\frac{d}{dt}\left( D_2 L_{(0,
1)}(t)\right)\right]{\mathbf \delta} y_{0}(t)\, dt\\
&& + \int^T_0 \left[D_3 L_{(N-1, N)}(t)-\frac{d}{dt}\left( D_4
L_{(N-1, N)}(t)\right)\right]{\mathbf \delta} y_{N}(t) \, dt\; .
\end{eqnarray*}

Therefore, the \emph{semi-discrete Euler-Lagrange equations} are:
\begin{eqnarray}
D_3 L_{(k-1, k)}(t) +D_1 L_{(k, k+1)}(t)- \frac{d}{dt} \left( D_4
L_{(k-1, k)}(t) + D_2 L_{(k, k+1)}(t)\right)&=&0,\quad 1\leq k\leq
N-1 \nonumber\\
D_1 L_{(0, 1)}(t)-\frac{d}{dt}\left( D_2 L_{(0,
1)}(t)\right)&=& 0 \label{ele1}\\
D_3 L_{(N-1, N)}(t)-\frac{d}{dt}\left( D_4 L_{(N-1,
N)}(t)\right)&=& 0 \nonumber
\end{eqnarray}

These are precisely the Euler-Lagrange equations
\[
\frac{\partial \widetilde{L}}{\partial
y_k}-\frac{d}{dt}\left(\frac{\partial \widetilde{L}}{\partial
\dot{y}_k}\right)=0, \quad 0\leq k\leq N
\]
for the Lagrangian $\widetilde{L}:
(TQ)^{N+1}\simeq TQ^{N+1}\longrightarrow \R$
defined by
\[
\widetilde{L}(y_0,\ldots, y_N, ({y}_0)_t, \ldots, ({y}_N)_t)
=\sum_{k=0}^{N-1} L( y_{k}, (y_{k})_t, y_{k+1}, (y_{k+1})_{t}),
\]
where  $(TQ)^{N+1}$ stands for the Cartesian
Product of $N+1$ copies of $TQ$.

We define
\begin{itemize}
\item The \emph{Poincar\'e Cartan 1-form} $\Theta_{\tilde{L}}\in
\Lambda^1(TQ^{N+1})$:
\[
\Theta_{\tilde{L}}=D_2L_{(0,1)}dy_0+\sum_{k=1}^{N-1}\left[D_4L_{(k-1,
k)}+D_2L_{(k, k+1)}\right] dy_k+D_4 L_{(N-1, N)}dy_N
\]
\item The \emph{Poincar\'e-Cartan 2-form} $\Omega_{\tilde{L}}=d\Theta_{\tilde{L}}$.
\item The \emph{energy function}
\[
E_{\tilde{L}}=D_2L_{(0,1)}(y_0)_t+\sum_{k=1}^{N-1}\left[D_4L_{(k-1,
k)}+D_2L_{(k, k+1)}\right] ({y}_k)_t+D_4 L_{(N-1, N)}
({y}_N)_t-\tilde{L}
\]
\end{itemize}
We say that the system is \emph{regular} if and only if the 2-form
$\Omega_{\tilde{L}}$ is a symplectic 2-form. Locally the
regularity is equivalent to the non-singularity of the Hessian
matrix $\displaystyle{\left(\frac{\partial^2 \tilde{L}}{\partial
(y_k)_{t}\partial (y_l)_{t}}\right)}$ $0\leq k,l\leq N$. In such a
case, there exists a unique vector field $\xi_{\tilde{L}}\in
{\frak X}(T(Q)^{N+1})$ such that
\begin{equation}\label{eqm}
i_{\xi_{\tilde{L}}}\Omega_{\tilde{L}}=dE_{\tilde{L}}
\end{equation}
Moreover, the integral curves of
$\xi_{\tilde{L}}$ are the tangent lifts of the
solutions of the semi-discrete Euler Lagrange
equations (\ref{ele1}).

In many situations, the 2-form $\Omega_{\tilde
L}$ is not symplectic. Then, (\ref{eqm}) has no
solution, in general, and even if it exists it
will not be unique. Let ${b}_{\tilde{L}}:
{TQ^{N+1}}\longrightarrow {T^*Q^{N+1}}$ be the
map defined by $b_{\tilde L}(X) = i_X
\Omega_{\tilde{L}}$. It may happen that
$\Omega_{\tilde{L}}$ is not surjective.
We denote by $\ker\Omega_{\tilde{L}}$ the kernel of $b_{\tilde
L}$, i.e., $\ker b_{\tilde L} = \ker\Omega_{\tilde{L}}$.

In \cite{GN1,GN2}, Gotay and Nester  have developed a constraint
algorithm for presymplectic systems which is an intrinsic version
of the classical Dirac-Bergmann algorithm . They consider the set
of points $P_2$ of $P_1=TQ^{N+1}$ where (\ref{eqm}) has a solution
and suppose that this set $P_2$ is a submanifold of $P_1$.
Nevertheless, these solutions on $P_2$ may not be tangent to
$P_{2}$. Then, we have to restrict $P_2$ to a submanifold where
the solutions of (\ref{eqm}) are tangent to $P_2$. Proceeding
further we obtain a sequence of submanifolds:
\[
\cdots\rightarrow P_k\rightarrow\cdots\rightarrow
P_2\rightarrow P_1=TQ^{N+1} \; .
\]

Alternatively, these constraint submanifolds may be described as
follows:
\[
P_i=\{p\in P_{i-1}\; / \; dE_{\tilde{L}}(p)(v)=0\; , \ \forall
v\in T_p P_{i-1}^{\perp} \;  \}\;  ,
\]
where
\[
T_p P_{i-1}^{\perp} = \{ v \in T_pP_1 \; / \;
\Omega_{\tilde{L}}(x)(u,v) = 0 \; ,\ \forall u \in T_p P_{i-1} \;
\} \; .
\]
We call $P_2$ the \emph{secondary constraint submanifold}, $P_3$
the \emph{tertiary constraint submanifold}, and, in general, $P_i$
is the $i$-\emph{ary constraint submanifold}.

If the algorithm stabilizes, i.e. there exists a positive integer
$k \in \N$ such that $P_k = P_{k+1}$ and $\dim P_k\not=0$, then we
have a \emph{final constraint submanifold} $P_f = P_{k}$, on which
exists a vector field $X$ such that
\begin{equation}\label{mf}
\left(i_X\Omega_{\tilde{L}}=dE_{\tilde{L}}\right)_{/P_f}
\; .
\end{equation}

If $\xi$ is a solution of (\ref{mf}) then every arbitrary solution
on $P_f$ is of the form $\xi' = \xi + Y$, where $Y \in
(\ker\Omega_{\tilde{L}}\cap TP_f)$.

Another interesting aspect of this theory is
that, in any case, regular or singular, since the
Lagrangian $\tilde{L}$ is autonomous, then
$\xi(E_{\tilde L})=0$ for any solution of
Equation (\ref{mf}). In particular, for
$\xi_{\tilde L}$, we have that
\[
\frac{d}{dt}\left[D_2L_{(0,1)}\dot
y_0+\sum_{k=1}^{N-1}\left[D_4L_{(k-1, k)}+D_2L_{(k, k+1)}\right]
\dot{y}_k+D_4 L_{(N-1, N)} \dot{y}_N-\tilde{L}\right]=0
\]
and the energy $E_{\tilde{L}}$ is a constant of the motion.

\subsubsection{Legendre transformation}

Define the semi-discrete Legendre  transformation
as
\[
\begin{array}{rrcl}
 \F\tilde{L}:& TQ^{N+1} &\longrightarrow &
 T^*Q^{N+1}\\
&(y_{0}, \ldots, y_{N},({y}_0)_t,\ldots,
({y}_N)_t) &\longmapsto& (y_0,\ldots, y_N; p_0,
\ldots, p_N)
\end{array}
\]
where
\[
p_0=D_2L_{(0,1)}, \ldots, p_k=D_4L_{(k-1, k)}+D_2L_{(k, k+1)},
\ldots, p_N=D_4 L_{(N-1, N)}
\]

If $\lambda_{Q^{N+1}}$ and $\omega_{Q^{N+1}}$ are the Liouville
$1$-form and the canonical symplectic 2-form on $T^*Q^{N+1}$ then
 \[
 \F\tilde{L}^*\lambda_{Q^{N+1}}=\Theta_{\widetilde{L}}, \quad
\F\tilde{L}^*\omega_{Q^{N+1}}=\Omega_{\widetilde{L}}
\]
Moreover, if the Legendre transformation $\F\tilde{L}$ is a global
diffeomorphism, we will say that the Lagrangian $\tilde{L}$ is
hyperregular. In this case it is well defined the function
$H=E_{\tilde{L}}\circ (\F\tilde{L})^{-1}$.

Therefore, we have a Hamiltonian representation of the equations
of motion (\ref{ele1}):
\[
i_{X_H}\omega_{Q^{N+1}}=dH
\]
or, in coordinates,
\[
\dot{y}_i=\frac{\partial H}{\partial p_i},\quad
\dot{p}_i=-\frac{\partial H}{\partial y^i}, \quad 0\leq i\leq N
\]
Obviously, $\xi_{\tilde{L}}$ and $X_H$ are $\F\tilde{L}$-related,
i.e. $(\F\tilde{L})_*\xi_{\tilde{L}}=X_H$.

If the lagrangian $\tilde{L}$ is singular, $\F\tilde{L}$ is not a
diffeomorphism. However, we may assume that $\tilde{L}$ is \emph{
almost regular}, i.e., $M_1=\F\tilde{L}(TQ^{N+1})$ is a
submanifold  of $T^*Q^{N+1}$ and, $\F\tilde{L}$ is a submersion
onto $M_{1}$ with connected fibers. The submanifold $M_1$ will be
called the \emph{primary constraint submanifold}.

Since the Lagrangian is almost regular, the
energy $E_{\tilde L}$ is constant along the
fibers of $\F\tilde{L}$. Therefore, $E_{\tilde
L}$ projects onto a function $H_{M_1}$ on $M_1$:
\[
H_{M_1}(\F\tilde{L}(p))=E_{\tilde L}(p)\; ,\forall p\in TQ^{N+1}
\; .
\]

If we denote by $\map{i_{M_1}}{M_1}{T^*Q^{N+1}}$
the embedding of $M_1$ into $T^*Q^{N+1}$, then we
obtain a presymplectic system $(M_1, (i_{M_1})^*
\omega_{Q^{N+1}}, dH_{M_1})$. If we now apply the
constraint algorithm to it, we shall obtain the
following sequence of constraint submanifolds:
\[
\cdots\rightarrow M_k\rightarrow\cdots\rightarrow M_2\rightarrow
M_1 \; ,
\]
as in the Lagrangian side.
 Denote by $M_f$ the final constraint
submanifold (if it exists) for this presymplectic system. In $M_f$
there  exists at least a  vector field $X\in {\frak X}(M_f)$ such
that
\[
\left(i_{X}\left[(i_{M_1})^*\omega_{Q^{N+1}}\right]=dH_{M_1}\right)_{/M_f}
\]

The Legendre map restricts to each submanifold $P_i$, $i\ge 1$, of
$TQ^{N+1}$ and then we obtain a family of surjective submersions
$({\F{\tilde L}})_i:{P_i}\longrightarrow {M_i}$ which relates the
constraint submanifolds $P_i$ and $M_i$, in particular, $P_f$ and
$M_f$.

\subsubsection{Momentum preservation}

Suppose, as in Subsection \ref{mp},  that the Lagrangian $L$ is
invariant by a Lie group of symmetries. Infinitesimally, this
conditions implies that
\[
\xi_{TQ^{N+1}}(v_{q_0},\ldots, v_{q_N})(\widetilde{L})=0, \quad
\forall \xi\in{\frak g}
\]
In such a case, applying the classical Noether theorem, we deduce
that  the function $F: TQ^{N+1}\longrightarrow \R$ defined by
\[
F=\Theta_{\tilde L}(\xi_{TQ^{N+1}})
\]
is a constant of the motion. Explicitly, \begin{eqnarray*}
 &&F(y_0,
\dot{y}_0, \ldots, y_N,
\dot{y}_N)=D_2L_{(0,1)}\xi_{Q}(y_0)\\
&&+\sum_{k=1}^{N-1}\left[D_4L_{(k-1, k)}+D_2L_{(k, k+1)}\right]
\xi_{Q}(y_k)+D_4 L_{(N-1, N)} \xi_{Q}(y_N) \end{eqnarray*}

\subsubsection{The nonlinear wave equation. Second point of view}

Consider the semidiscrete  Lagrangian
\[
 L_{sd}(u_0, u_1, (u_0)_t, (u_1)_t)= \frac{1}{2}
\left(\frac{(u_0)_t+(u_1)_t}{2}\right)^2
-\sigma\left(\frac{u_1-u_0}{h}\right)-f\left(\frac{u_1+u_0}{2}\right)
\]
for the nonlinear wave equation. Then the Euler-Lagrange equations
are: \begin{eqnarray*}
(a)&&\frac{1}{4}\left(({u}_{0})_{tt}+(u_1)_{tt}\right)-\frac{1}{h}\left[\sigma'\left(\frac{u_{1}-u_{0}}{h}\right)\right]+\frac{1}{2}f'(\frac{u_0+u_{1}}{2})=0\\
(b)&&\frac{({u}_{k-1})_{tt}+2({u}_k)_{tt}+({u}_{k+1})_{tt}}{4}-\frac{1}{h}\left[\sigma'\left(\frac{u_{k+1}-u_k}{h}\right)
-\sigma'\left(\frac{u_{k}-u_{k-1}}{h}\right)\right]\\
&&+\frac{1}{2}\left[f'(\frac{u_k+u_{k-1}}{2})
+f'(\frac{u_{k+1}+u_{k}}{2})\right]=0,\quad 1\leq k\leq N-1\\
(c)&&\frac{1}{4}\left(({u}_{N})_{tt}+({u}_{N-1})_{tt}\right)+\frac{1}{h}\left[\sigma'\left(\frac{u_{N}-u_{N-1}}{h}\right)\right]+\frac{1}{2}f'(\frac{u_{N-1}
+u_{N}}{2})=0.
\end{eqnarray*}

The Lagrangian ${\tilde L}$ is
\begin{eqnarray*}
 \tilde{L}(u_0,\ldots, u_N; ({u}_0)_t, \ldots,
({u}_N)_t)&=&\sum_{k=1}^N \left[\frac{1}{2}
\left(\frac{(u_{k-1})_t+(u_k)_t}{2}\right)^2\right.\\
&&\left.-\sigma\left(\frac{u_k-u_{k-1}}{h}\right)-f\left(\frac{u_k+u_{k-1}}{2}\right)\right]
\end{eqnarray*}
and the  Poincar\'e-Cartan 1-form is
\begin{eqnarray*}
\Theta_{\tilde L}&=&\frac{1}{4}\Big[
\left((u_0)_t+(u_1)_t\right)du_0 +
\sum_{k=1}^{N-1}\left((u_{k-1})_t+2(u_k)_t+(u_{k+1})_t\right)du_k\\
&& +\left((u_{N-1})_t+(u_{N})_t\right)du_N\Big]
\end{eqnarray*}

In this case the Poincar\'e Cartan 2-form $\Omega_{\tilde
L}=-d\Theta_{\tilde L}$ is degenerate with
\[
\ker \Omega_{\tilde
L}=\hbox{span}\left\{\sum_{k=0}^N (-1)^k
\frac{\partial}{\partial u_k}, \sum_{k=0}^N
(-1)^k \frac{\partial}{\partial (u_k)_t} \right\}
\]
and, therefore, it is necessary to implement the constraint
algorithm. We will study in a future paper the constraint
algorithm for this and other examples.

\section{Conclusions and outlook}
In this paper we have elucidated the geometrical
framework for semi-discrete Mechanics, an useful
tool for numerical simulation of Classical Field
theories. For simplicity, we  only have studied
the case of the lagrangian $L_{sd}: TQ\times
TQ\simeq T (Q\times Q) \longrightarrow \R$ but
also it is possible to consider extensions of
this theory to the case of Lagrangians $L_{sd}:
{\mathcal P}^{\tau}G \longrightarrow \R$ where
${\mathcal P}^{\tau}G$ is the prolongation of a
Lie groupoid $G\rightrightarrows M$    over the
fibration $\tau: AG\longrightarrow \R$, $AG$
being the Lie algebroid associated to the Lie
groupoid $G$ (see \cite{groupoid} for more
details). This extension may be useful for
generating numerical schemes for  Field Theories
modeled on Lie algebroids (see \cite{mart2}).

Moreover, the semi-discrete lagrangian $L_{sd}: TQ\times TQ
\longrightarrow \R$ is adapted for field theories whose lagrangian
$L: Z\longrightarrow \R$ does not depend on the base variables
(the base space is assumed bi-dimensional) , that is, $L=L(y^i,
z^i_{\mu})$. Suppose, for simplicity, that the parameter space is
$X=\R\times P$, with $P=\R^2$ and $Y=X\times \R$, and the
continuous Lagrangian is of the form
\[
L=L(t, x^1, x^2,  y, z_{t}, z_{x_1}, z_{x_2})
\]
Introduce the following natural semi-discretization of $L$
\begin{eqnarray*}
&&(L_{sd})_{(k_1, k_2)}(t, y_{(k_1, k_2)}(t),y_{(k_1+1,
k_2)}(t),y_{(k_1, k_2+1)}(t); \dot{y}_{(k_1,
k_2)}(t),\dot{y}_{(k_1+1, k_2)}(t),\dot{y}_{(k_1, k_2+1)}(t))
\\
&&=L(t, k_1h_1, k_2 h_2, \frac{y_{(k_1, k_2)}(t)+y_{(k_1+1,
k_2)}(t)+y_{(k_1, k_2+1)}(t) }{3},\\
&& \frac{\dot{y}_{(k_1, k_2)}(t)+\dot{y}_{(k_1+1,
k_2)}(t)+\dot{y}_{(k_1, k_2+1)}(t) }{3}, \frac{y_{(k_1+1,
k_2)}(t)-y_{(k_1, k_2)}(t)}{h_1}, \frac{y_{(k_1,
k_2+1)}(t)-y_{(k_1, k_2)}(t)}{h_2}) \end{eqnarray*} based on a
triangularization of the space $\Z\times \Z$. Then we need to
extremize a functional of the type
\begin{eqnarray*}
 {\mathcal
S}_{\hbox{\small sd}} L =\int^T_0
\Big[\sum_{(k_1, k_2)\in \{0,\ldots, N-1\}\times
\{0,\ldots, N-1\}}(L_{sd})_{(k_1, k_2)}(t,
y_{(k_1, k_2)}(t),y_{(k_1+1, k_2)}(t),y_{(k_1,
k_2+1)}(t);&&\\
 \dot{y}_{(k_1, k_2)}(t),\dot{y}_{(k_1+1,
k_2)}(t),\dot{y}_{(k_1, k_2+1)}(t))\Big]\, dt&&
\end{eqnarray*}
Observe that $(L_{sd})_{(k_1, k_2)}\in
C^{2}(\R\times T\R^3)$ and then new tools must be
used as, for instance, cosymplectic geometry and
suitable adaptations of the results about
Discrete Field Theories (see, for instance,
\cite{mars,joris}). Also extensions for the case
of a non-uniform grid may  be considered and in a
future paper will be discussed. The
generalization of this theory to the case of
constrained multisymplectic field theories, as
for instance in the case of incompressibility
constraints in fluids \cite{MM}, may be obtained
using and adaptation of constrained discrete
variational calculus (see \cite{BM}, for
instance).

\end{document}